\documentclass[prl,aps,twocolumn,showpacs,superscriptaddress]{revtex4}

\usepackage{graphicx}
\usepackage{epsfig}

\newcommand{\rmd}{{\rm d}}
\newcommand{\FF}{\mathcal{F}}

\newcommand{\ket}[1]{|#1\rangle}
\newcommand{\beq}{\begin{equation}}
\newcommand{\eeq}{\end{equation}}

\begin{document}
\title{Defect-Suppressed Atomic Crystals in an Optical Lattice}
\author{P. Rabl}
\author{A. J. Daley}
\author{P. O. Fedichev}
\affiliation{Institut f\"{u}r Theoretische Physik, Universit\"{a}t Innsbruck, A-6020 Innsbruck, Austria}
\author{J. I. Cirac}
\affiliation{Max-Planck Institut f\"{u}r Quantenoptik, D-85748 Garching, Germany}
\author{P. Zoller}
\affiliation{Institut f\"{u}r Theoretische Physik, Universit\"{a}t Innsbruck, A-6020 Innsbruck, Austria}
\date{1 April 2003}

\begin{abstract}
We present a coherent filtering scheme which dramatically reduces the site occupation number defects for atoms in an optical lattice, by transferring a chosen number of atoms to a different internal state via adiabatic passage. With the addition of superlattices it is possible to engineer states with a specific number of atoms per site  (atomic crystals), which are required for quantum computation and the realisation of models from condensed matter physics, including doping and spatial patterns. The same techniques can be used to measure two-body spatial correlation functions. 
\end{abstract}

\maketitle

There has been a vast amount of recent interest in the study of Bose-Einstein Condensates (BECs) \cite{naturereview} and degenerate Fermi gases \cite{fermiexp} in optical lattices \cite{bosonsol,lattice,blochsdol}. Such systems have many potential applications in quantum computing, and also make possible the study of strongly correlated systems from condensed matter theory with unprecedented control over system parameters.

However, the study of strongly correlated systems or entangled atoms in optical lattices requires not only that the 
corresponding (Hubbard) Hamiltonian \cite{bosonsol,lattice} be properly engineered, but also that the system is placed in a well defined (pure) initial state with high precision. Whilst is possible to load a BEC into an optical lattice in the Mott Insulator (MI) phase, which corresponds to having a definite number of atoms at each lattice site \cite{bosonsol,lattice}, non-ideal conditions will always result in defects in that phase (i.e., missing atoms and overloaded sites). The removal (or precise control) of such defects is then a necessary condition for the application of atoms in optical lattices to quantum computing and the study of strongly correlated systems.
In this letter, we propose a coherent filtering scheme which, beginning with an uncertain number of atoms in each site, provides a method to transfer a definite number of atoms at each site into a different internal state, and so load a new lattice of atoms with an exact number of particles per site. This process dramatically reduces the site 
occupation number defects, and can be extended, under experimentally reasonable conditions, to allow the production of doped or pattern loaded initial states with almost unit fidelity (i.e., atomic crystals). The scheme is also applicable to the production of high fidelity initial states in fermion systems (including doped states and the loading of composite objects), and when combined with moving optical lattices, can be used to measure two-body correlation functions. We illustrate this by presenting an example of how this scheme could be used to both produce and characterise superconducting states in an optical lattice with a chosen filling factor. 

\begin{figure}[t]
\begin{center}
\includegraphics[width=7cm]{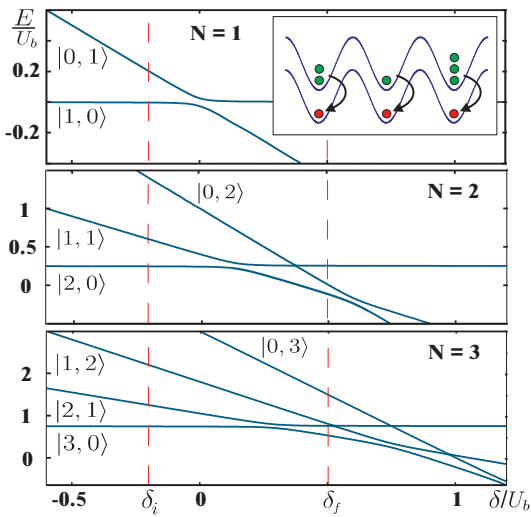}
\end{center}
\caption{Plots showing avoided crossings in the energy eigenvalues for $N=1,2,3$ (Note the vertical scale variation). We must choose $\delta_i$ and $\delta_f$ (dashed vertical lines) so that we cross only
from $\ket{N,0}\rightarrow\ket{N-1,1}$, as illustrated in the inset.} 
\label{crossings}
\end{figure}

We begin by considering a system of bosons loaded into an optical lattice such that they do not tunnel between neighbouring sites. The atoms are in a particular internal state, $\ket{a}$, and have onsite interaction strength (with $\hbar=k_B=1$) $U_a=4\pi a_s \int \rmd^3 {\mathbf x}|{\rm w}({\mathbf x})|^4 /m$, (valid for $U_a$ less than the separation of the two lowest motional states at each site) where $a_s$ is the scattering length and ${\rm w}({\mathbf x})$ is a Wannier function \cite{bosonsol}. We then couple the atoms into a second internal state, $\ket{b}$, (which is trapped by a second lattice potential) via an off-resonant Raman transition with Rabi frequency $\Omega(t)$, which is detuned from state $\ket{b}$ by $\delta(t)$. If we denote the onsite interaction between particles in state $\ket{b}$ as $U_b$ and the onsite interaction between particles in different internal states as $U_{ab}$, and follow the derivation in \cite{bosonsol}, we obtain the Hamiltonian
\begin{eqnarray}
\hat H &=& \,\frac{U_{a}}{2} \hat n_{a} (\hat n_{a} - 1) + \frac{U_{b}}{2} \hat n_{b} (\hat n_{b} - 1) + \,U_{ab} \hat n_{a} \hat n_{b}\nonumber\\
 & & -\frac{\Omega(t)}{2} ( a^{\dagger} b + b^{\dagger} a)- \delta(t) \hat n_{b},\label{mainhamiltonian}
\end{eqnarray}
where $\hat a$ and $\hat b$ are the annihilation operators for particles in states $\ket{a}$ and $\ket{b}$ respectively, and $\hat n_a=\hat{a}^\dag \hat a$, $\hat n_b=\hat{b}^\dag \hat b$. We can then write the state of the site as $\ket{n_a,n_b}$, where $n_a$ and $n_b$ are the number of particles in the internal states $\ket{a}$ and $\ket{b}$ respectively. 
We know that the initial state at a particular lattice site is $\ket{N,0}$, where $N \in {1,2,\ldots,N_{\rm max}}$. Our goal is to transfer exactly one particle from state
$\ket{a}$ to state $\ket{b}$, so that for any $N$, the final state is $\ket{N-1,1}$. In order to do this we must choose the initial and final values of the detunings, $\delta_i$ and $\delta_f$ so that the system evolves along the avoided
crossing in the energy eigenvalues, undergoing an adiabatic passage from $\ket{N,0}\rightarrow\ket{N-1,1}$, and does not evolve along any other such avoided crossings. These values have to be simultaneously chosen for all values of $N$, as shown in Fig.~\ref{crossings}. Once the system is in state $\ket{N-1,1}$, we can turn off the lattice trapping state $\ket{a}$, leaving a pure state with exactly one atom per lattice site.

\begin{figure}[hbt]
\begin{center}
\includegraphics[width=8cm]{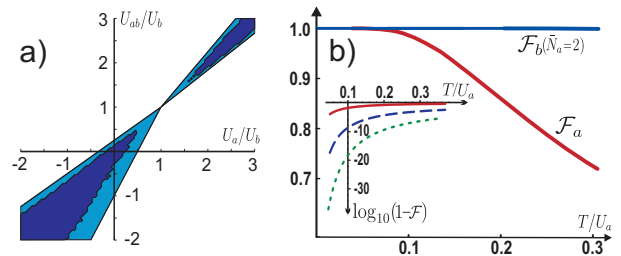}
\end{center}
\caption{a) The parameter range for which appropriate values of $\delta_i$ and $\delta_f$ can be found when $N_{\rm max}=4$, in the adiabatic limit (light shading) and from numerical simulations with a smoothed rectangular pulse which rises and falls with a $\sin^2(t)$ shape, $\tau=100 U_b^{-1}$ and max $\Omega=0.3 U_b$, giving a transfer error $\varepsilon<1$\% (dark shading). b) Initial and final state fidelities, $\FF_a$ and $\FF_b$, as a function of temperature, $T$, for an initial MI state described at each site using the GCE. The inset shows values of $1-\FF_a$ (solid line) and $1-\FF_b$ for $\overline{N}_a=2$ (dashed) and $\overline{N}_a=3$ (dotted) on a logarithmic scale.} 
\label{param_fluct}
\end{figure}

The relative locations of the avoided crossings in the energy eigenvalues are determined by the values of $U_{a}/U_{b}$ and $U_{ab}/U_b$. In
the adiabatic limit ($\dot{\delta}\rightarrow 0$), we can set $\Omega=0$, and for a given $N_{\rm max}$ it is then straight forward to find the parameter range for which an appropriate choice of $\delta_i$ and $\delta_f$ is
possible. We consider the locations at which the energy eigenvalues of $\hat H$ cross in this limit, and then attempt to find a parameter range for which
the crossings between $\ket{N,0}$ and $\ket{N-1,1}$ for all $N$ fall within a range of detunings $[\delta_i,\delta_f]$ which isolates them from all other crossings. The resulting boundaries for the allowed parameter space are given by
$U_{ab}=(1-k_i) U_a + k_i U_b$, for $i=1,2$, where
for $U_{ab}>U_a$, $k_1=1/N_{\rm max}$, $k_2=1/(2 N_{\rm max}-4)$, and for $U_{ab}<U_a$, $k_1=1/(2 N_{\rm max}-2)$, $k_2=1/(N_{\rm max}-3)$.
These boundaries are shown in Fig. \ref{param_fluct}a for $N_{\rm max}=4$. Larger values of $N_{\rm max}$ result in a more restrictive allowed parameter range. Substantially different values of $U_a$ and $U_b$ are required, which is possible using either spin-dependent lattices \cite{blochsdol}, where $U_a$ and $U_b$ are independently controlled via the different lattice shape for atoms in each internal state, or Feshbach Resonances \cite{bosefeshbach}, near which the scattering length $a_s$ is different for atoms in different internal states as we tune an external magnetic field. $U_{ab}$ can also be independently controlled by slightly displacing the lattices trapping internal states $\ket{a}$ and $\ket{b}$.

In the case of finite-time laser pulses, the adiabatic crossings have a finite size, and must be crossed on a timescale dictated by the separation of the energy eigenvalues at the crossings. From the Landau-Zener formula, we can estimate the transfer error, $\varepsilon$, for $\sqrt{N_{\rm max}}\Omega(t)<U_a,U_b$ as $\varepsilon\approx \exp[-\pi N\Omega^2/(2 \dot{\delta})]$, where $\dot{\delta}\sim |\delta_f-\delta_i|/\tau$ and $\tau$ is the total transfer time. For a large range of parameter values, we find that $\varepsilon$ is very small for $\tau$ values well within the decoherence time (including spontaneous emissions, collisional losses, and external parameter fluctuations, \cite{lattice}) of the optical lattice. This has been confirmed by extensive numerical calculations of the exact time evolution of (\ref{mainhamiltonian}) (Fig. \ref{param_fluct}a).

This filtering scheme naturally cannot correct occupation number defects in which no atoms are present at particular lattice site in state $\ket{a}$. Because in principle extremely low transfer errors, $\varepsilon$, are obtainable, this is the fundamental limit for the filtering scheme. The initial single-site fidelity of a state where we choose to have $N_a$ particles in each lattice site is defined as $\FF_a=\langle N_a,0|\hat w_0|N_a,0\rangle$, where $\hat w_0$ is the initial density operator for the site. Similarly, the final state single-site fidelity (for 1 transferred atom) is $\FF_b=\langle 1|\hat w_b|1\rangle$, where $\hat w_b={\rm Tr}_a \hat w$, and $\hat w$ is the final density operator for the site. In each case the overall fidelity for $M$ sites is given by $\FF_M=\FF^M$. If the probability of zero occupation in a particular site is $p_0$, then $\FF_b=(1-\varepsilon)(1-p_0)$. Thus to obtain the highest possible fidelities we should start in a MI phase with an average filling factor of 2 or more particles per site, where $p_0$ is small. For such a MI phase, with an average filling factor of $\overline{N}_a$ at temperature $T$, we model the particle number fluctuations using a Grand Canonical Ensemble (GCE). For integer $\overline{N}_a$ and low initial temperature, $\exp[-U_a/(2T)]\ll 1$, we obtain $p_0=\exp[-U_a \overline{N}_a^2/(2T)]$, so that for $\varepsilon \rightarrow 0$, $\FF_b\approx (1-\exp[-U_a \overline{N}_a^2/(2T)])$. Under the same conditions, we can write $\FF_a=p(N=\overline{N}_a)\approx 1-2\exp[-U_a/(2T)]$. We see (Fig. \ref{param_fluct}b) that coherent filtering increases the fidelity by several orders of magnitude, even when the fidelity $\FF_a$ is already reasonably high. Moreover, as $(1-F_b)=[(1-F_a)/2]^{\overline{N}_a}$, we see that the error in the initial state is exponentially suppressed with exponent $\overline{N}_a$. For example, if we assume $\varepsilon\rightarrow 0$, an initial state with $\overline{N}_a=2$ and a defect at every tenth lattice site, $\FF_a=0.9$, $T/U_a=0.17$, results in a final state with $\FF_b\approx 3\times 10^{-6}$, i.e., less than one defect every three hundred thousand sites. In an attempt to reduce number fluctuations even further we also considered a regime in which the atoms were allowed to move between neighbouring sites during the filtering process. However, for a fixed mean filling factor and transfer time this produces a lower final state fidelity than performing the filtering in the MI regime.  

Taking Poisson statistics instead of the GCE, which corresponds to the BEC being loaded by suddenly switching on the lattice (this can be seen by computing the reduced density operator of the original superfluid state), and assuming a perfect transfer ($\varepsilon=0$), we obtain $\FF_b=1-\exp(-\overline{N}_a)$. Thus, to achieve high fidelities $\FF>0.99$ in the final state requires $\bar{N}_a\sim 5$. Such $\overline{N}_a$ values might ultimately be limited by the increase in 3-body collisional losses with many atoms on each site.

This filtering can clearly be extended to transfer multiple atoms into state $\ket{b}$ by choosing $\delta_f$ so that the system evolves along more than one avoided crossing (but the same number for all initial $N$). Similarly, it is possible to choose $\delta_i$ and $\delta_f$ so that the transfer $\ket{N,0}\rightarrow\ket{N-1,1}$ occurs only for initial states with a particular value of $N$. Thus, we can measure the defects in the initial state, for example by mapping sites with $N_a=\overline{N}_a+1$ or $N=\overline{N}_a -1$ atoms onto occupied sites in state $\ket{b}$. 

Spatial patterns can be created by adding a superlattice to the lattice trapping atoms in state $\ket{b}$, site-dependently shifting the energy level of $\ket{b}$ and preventing atoms in particular sites from coupling to the Raman transition. This allows the engineering of many high-fidelity spatial states because of the wide range of possible superlattice configurations which can be formed using overlapping laser beams.

\begin{figure}[htb]
\begin{center}
\includegraphics[width=8cm]{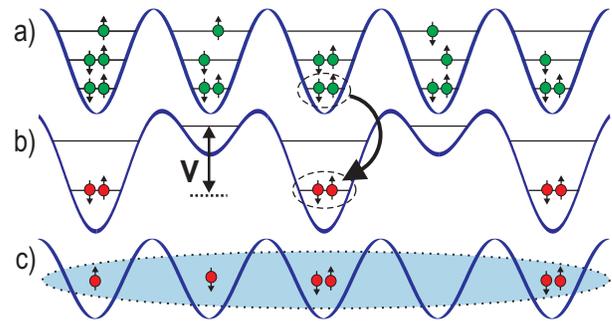}
\end{center}
\caption{The production of (c) a BCS state at exactly half filling in an optical lattice from (a) a degenerate Fermi gas using pattern loading techniques described in the text (a$\rightarrow$b).} 
\label{bcsscheme}
\end{figure}

Coherent filtering can also be applied to a degenerate Fermi gas \cite{fermiexp, fermifeshbach} in an optical lattice (Fig. \ref{bcsscheme}a). We consider spin up and spin down fermions (in states $\ket{a\uparrow,\downarrow}$) loaded into the motional levels of each lattice site according to the Pauli principle. Atoms in any chosen motional level (normally the lowest, as for low temperature and high filling factors it will almost always be occupied) may then be transferred by adiabatic passage to a different internal state, $\ket{b\uparrow,\downarrow}$. An example of appropriate states can be taken from the $2$S$_{1/2}$ level of $^6$Li in a strong magnetic field (i.e., in the Paschen-Back regime). We choose $\ket{a\uparrow,\downarrow}=\ket{{\rm m_s=\frac{1}{2};m_I=0,-1}}$ and $\ket{b\uparrow,\downarrow}=\ket{{\rm m_s=-\frac{1}{2};m_I=1,0}}$, where ${\rm m_s}$ and ${\rm m_I}$ denote the magnetic spin quantum number for the electronic and nuclear spins respectively. (Where required, spin-dependent lattices for these states could be implemented for the short times required by coupling the ground states between the $2$P$_{1/2}$ and $2$P$_{3/2}$ levels.)
As was the case for Bosons, we can apply a superlattice to the lattice which traps state $\ket{b}$, and hence engineer very high fidelity spatial patterns (Fig.~\ref{bcsscheme}b). By applying different superlattices during the transfer of different spin states, the ratio of spin up to spin down particles in the final state can also be very precisely controlled. 

These techniques, when combined with the ability to move spin-dependent optical lattices \cite{blochsdol}, can be used to measure two body correlation functions. Currently, sinlge-body correlation functions can be studied using the interference patterns produced when atoms are released from the lattice, and many-body correlation functions for a single site can be estimated from collisional loss rates. However, for fermion systems in a lattice, fourth order correlation 
functions of the form $\rho_{i,j,l}=\langle \hat{c}^\dag_{i+l \uparrow}\hat{c}^\dag_{i-l \downarrow}\hat{c}_{j-l\downarrow} \hat{c}_{j+l\uparrow} \rangle$ 
(where $c_{i \sigma}$ is the annhilation operator for an atom of spin $\sigma$ in site $i$) are of particular interest because they provide information about both 
about the long range order and the pair correlation length in the system. Specifically, 
$\kappa_{l}=\sum_{i}\langle \hat{c}^\dag_{i+l \uparrow}\hat{c}^\dag_{i-l \downarrow}\hat{c}_{i-l \downarrow} \hat{c}_{i+l \uparrow} \rangle$ 
characterises the correlation as a function of the separation $2 l$ between spin up and spin down particles, and $\gamma_{k}=\sum_i\langle \hat{c}^\dag_{i-k\uparrow}\hat{c}^\dag_{i-k\downarrow}\hat{c}_{i+k \downarrow} \hat{c}_{i+k \uparrow} \rangle$ 
gives us information about the long range order as a function of the separation $2 k$ between the pairs. These two functions can be measured by firstly making the lattice potential very deep, and performing the measurement on a timescale shorter than that on which atoms can move to neighbouring sites. Then, to measure $\kappa_l$, we first transfer all of the spin down atoms from their current state, 
$\ket{b\downarrow}$ via an adiabatic passage to the internal state $\ket{a\downarrow}$. We then shift the lattice trapping the state $\ket{a\downarrow}$ $2l$ 
sites to the left. If we now transfer the particles in state $\ket{a\downarrow}$ to state $\ket{b\downarrow}$ selectively, based on the condition that no 
particle exists (at a particular site) in the state $\ket{b\uparrow}$, then the number of atoms left in state $\ket{a\downarrow}$ is proportional to the 
value of $\kappa_l$. This is made possible by the onsite interaction between particles with different spins shifting the detunings at which avoided crossings occur 
in the energy eigenvalues. A similar process can be used to measure $\gamma_{l}$, but requires an additional step in which we flip the spin of particles 
in state $\ket{b\downarrow}\rightarrow\ket{b\uparrow}$. Whilst these processes would be difficult to implement experimentally, they can be seen as 
providing an eventual way to perform a complete tomography of pair correlations in fermion systems. 

\begin{figure}[hbt]
\begin{center}
\includegraphics[width=4cm]{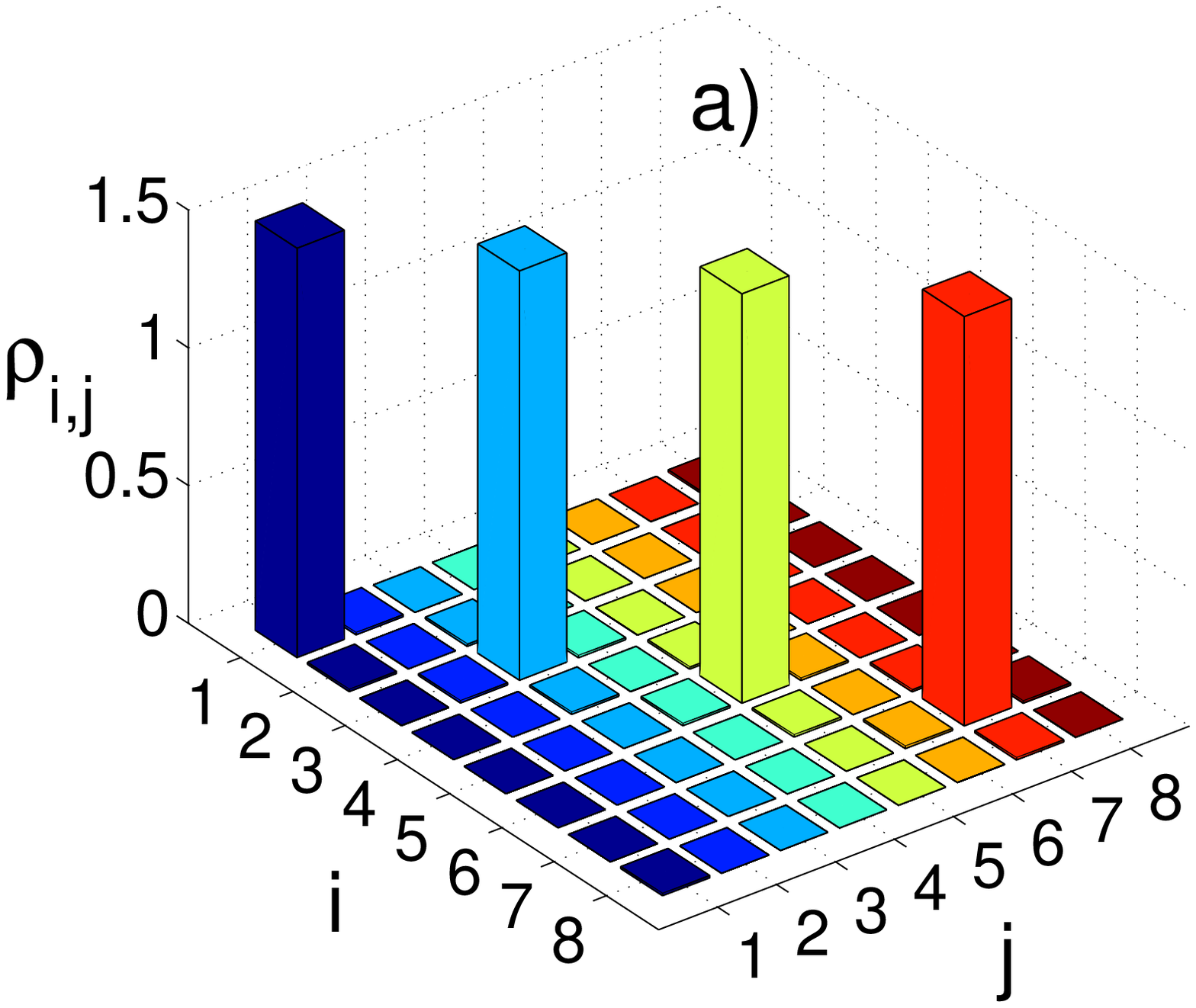}
\includegraphics[width=4cm]{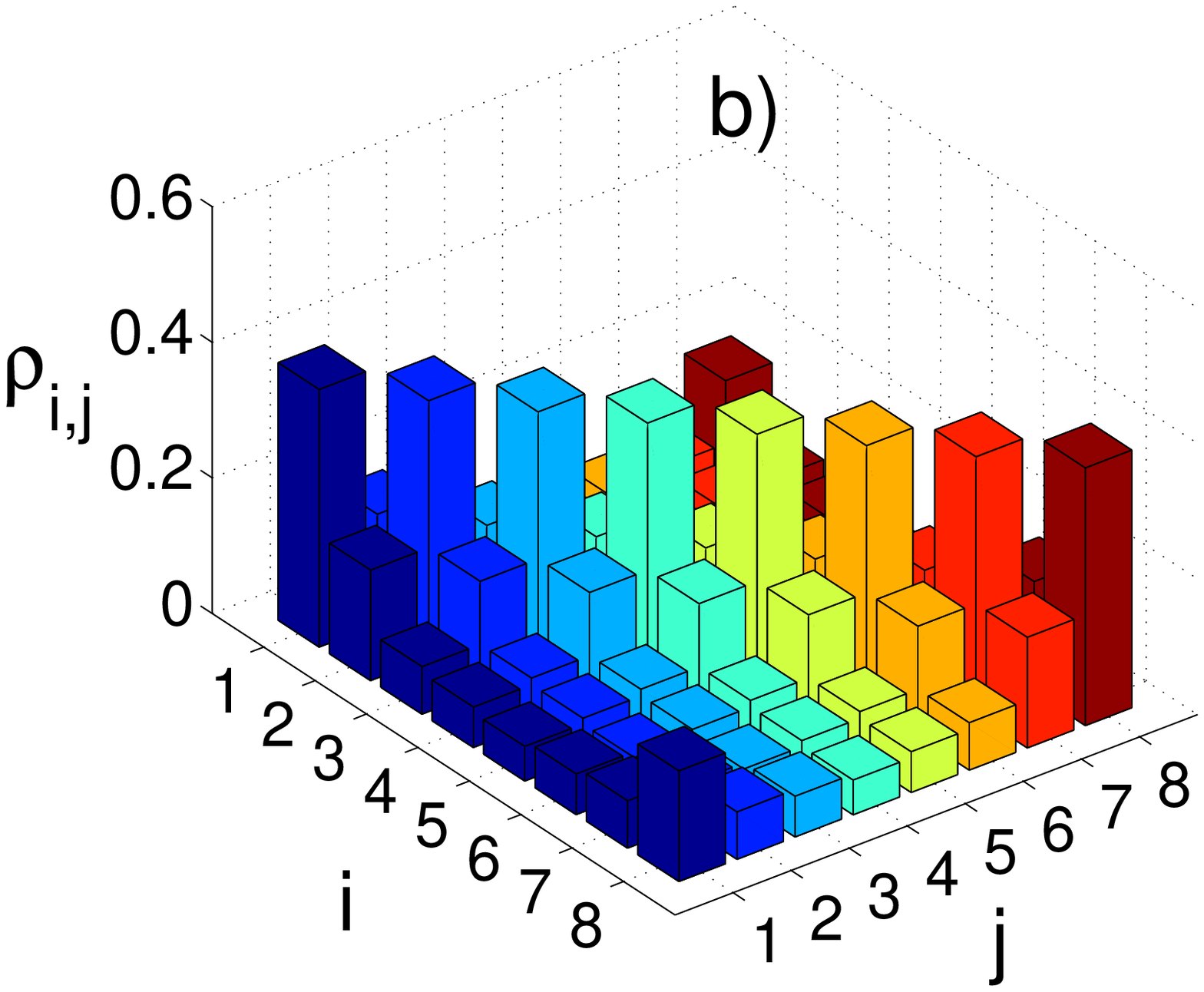}
\end{center}
\caption{The pair correlation function $\rho_{i,j,0}$, illustrating the crossover from MI$\rightarrow$BCS states. In (a) ($V=10J$, $U=-10J$, MI regime) and (b) ($V=0$, $U=-0.1J$, BCS regime) we see localized pairs and delocalised pairs respectively. Information about these correlations can be obtained from $\gamma_l=\sum_i \rho_{i-l,i+l}$. These results are from numerical diagonalisation of (\protect{\ref{bcsham}}) for 8 sites and half filling with periodic boundary conditions.} 
\label{lrocorr}
\end{figure}

\begin{figure}[htb]
\begin{center}
\includegraphics[width=4cm]{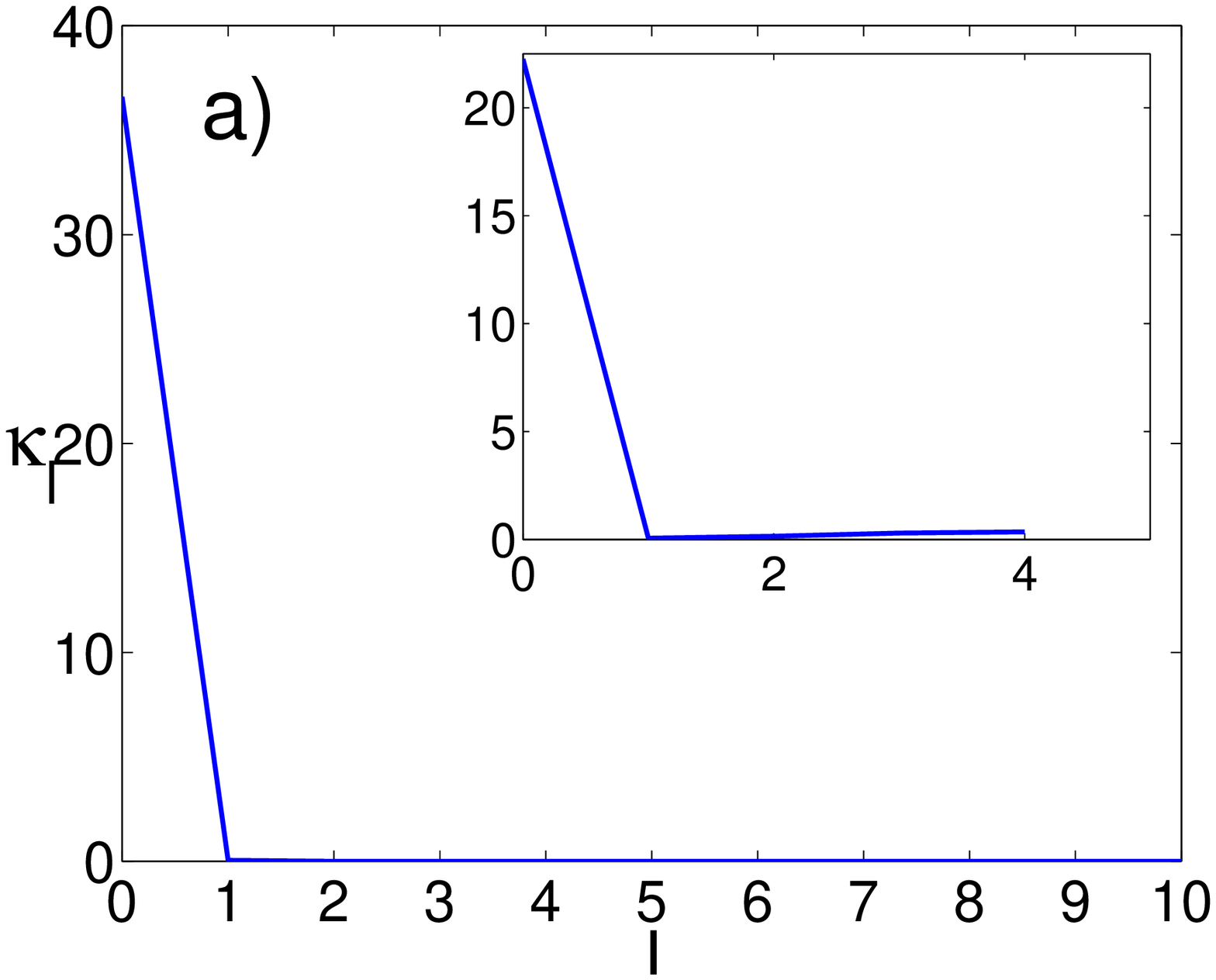}
\includegraphics[width=4cm]{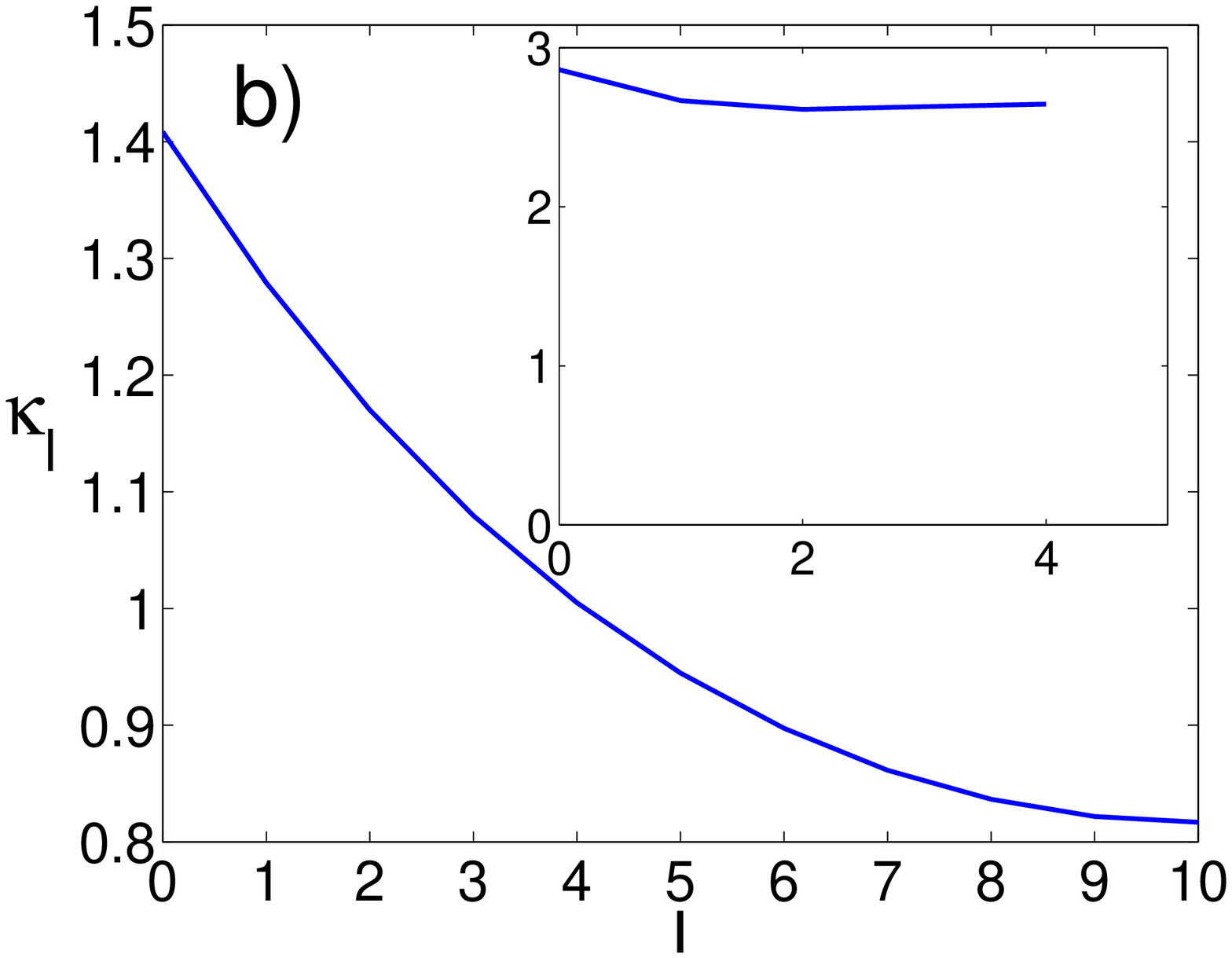}
\end{center}
\caption{Numerical values of $\kappa_{l}$, which describes the pair correlation length. These results are examples from numerical diagonalisation of (\protect{\ref{bcsham}}) for 40 sites and 1 particle of each spin type and 16 sites and 2 particles of each spin type (inset) with a) $V=0$, $U=-10 J$ and b) $V=0$, $U=-0.1 J$.} 
\label{iplcorr}
\end{figure}

Such a characterisation would be particularly useful in the study of the BEC - BCS crossover in an optical lattice \cite{supercondnarrowband}. Along with initial state preparation, this is well illustrated by the method to study the MI to BCS state transition in fermions at exactly half-filling shown in Fig. \ref{bcsscheme}. We begin by transfering particles of both spin types from a degenerate Fermi gas loaded into an optical lattice (Fig. \ref{bcsscheme}a) via adiabatic passage to a different internal state, whilst a superlattice is used to offset even-numbered sites in the final state by a potential $V$ (Fig. \ref{bcsscheme}b). This system is then described by the Hamiltonian
\beq
\hat{H}=-J\sum_{\langle i,j \rangle,\sigma}\hat{c}^\dag_{i\sigma}\hat{c}_{j\sigma} +U\sum_{i=1}^N \hat{n}_{i\uparrow}\hat{n}_{i\downarrow}+\frac{V}{2}\sum_{i,\sigma}^N (-1)^i \hat{n}_{i,\sigma},
\label{bcsham}
\eeq
where $\langle i,j \rangle$ denotes all nearest neighbour combinations and $\hat{n}_{i,\sigma}=\hat{c}^\dag_{i,\sigma}\hat{c}_{i,\sigma} $. $U$ denotes the onsite interaction
between particles of different spin, which we choose to make initially large and negative, using a combination of lattice parameters, 
and, if necessary, a Feshbach resonance \cite{fermifeshbach}.
Thus we prepare a state with exactly half-filling (note that other filling factors are possible with different superlattices), consisting of pre-formed pairs in the odd numbered sites (Fig. \ref{lrocorr}a).
We slowly decrease $V$ to a small value, $V<J \ll U$, allowing the pairs to delocalise. For small $V$ but large $|U|$, the pairs behave as a gas of hard-core Bosons. Following the discussion in \cite{zwergercommincomm}, we find that the system is protected throughout this process by an energy gap of $V$. This resulting state is characterised by a very short inter-particle correlation length, measureable through the correlation function $\kappa_l$ (Fig. \ref{iplcorr}a).

We then lower the onsite interaction until $|U|<J$, so that the spin up and spin down particles can move independently. A meanfield treatment \cite{daleyfedichevmf} shows that even in the weak coupling limit the state is protected by an energy gap which is no less than $V$. Once the BCS state of Cooper pairs sets in \cite{supercondnarrowband},  we can set $V\rightarrow 0$ by completely removing the superlattice. The system thus undergoes a transition into a BCS state (Fig. \ref{bcsscheme}c, \ref{lrocorr}b), characterised by its long pair correlation length, again measureable using $\kappa_l$ (Fig. \ref{iplcorr}b). 

In summary, the coherent filtering scheme presented here allows the production of high-fidelity atomic crystals, and the measurement of important correlation functions. These features will greatly enhance the application of atoms in optical lattices to quantum computation and to the precise modeling of condensed matter systems.

We thank D. Jaksch, L. Tian, M. Bijlsma, and members of R. Grimm's group for discussions.
Work was supported in part by the Austrian Science Foundation, E.U. Networks, and the Institute for Quantum Information.


\end{document}